\documentstyle[preprint,aps]{revtex}
\begin{document}

\draft
\title
{\bf Interdot Coulomb repulsion effect on the charge transport of
parallel double single electron transistors.}

% {\bf Single electron transistors as  highly sensitive charge
%detectors}

\author{David M.-T. Kuo and P. W. Li}
\address{Department of Electrical Engineering, National Central
University,\\
Chung-Li, Taiwan 320, Republic of China}

%\end{center}
\date{\today}
\maketitle

\begin{abstract}
The charge transport behaviors of parallel double single electron
transistors (SETs) are investigated by the Anderson model with two
impurity levels. The nonequilibrium Keldysh Green's technique is
used to calculate the current-voltage characteristics of system.
For SETs implemented by quantum dots (QDs) embedded into a thin
$SiO_2$ layer, the interdot Coulomb repulsion is more important
than the interdot electron hopping as a result of high potential
barrier height between QDs and $SiO_2$ barrier. We found that the
interdot Coulomb repulsion not only leads to new resonant levels
but also creates negative differential conductances.
\end{abstract}

\newpage

\section{Introduction}
The transport properties of a single electron transistor (SET)
have been experimentally and theoretically studied by many groups
[1-5]. Moreover, the interesting phenomena of SETs such as Kondon
effect and Coulomb blockade have been demonstrated
experimentally[1-3]. SETs comparing with conventional transistors
offer the advantages of good scalability, ultralow power operation
for future memory and quantum information applications. Recently,
some Si or Ge-dot SETs fabricated by CMOS-compatible process
techniques have been proposed and large Coulomb-blockade
oscillations at room temperature have also been reported[6-9].

For practical integrated-circuit applications, it is crucial to
clarify that if the current-voltage characteristics of a single
SET will be affected by the nearby SETs or quantum dots (QDs). The
main purpose of this work is to theoretically study the charge
transports of SETs sitting in parallel as depicted in Fig. 1,
where A-SET (with dot A) is regarded as a detector SET and B-SET
(with dot B) is as a target SET. Each SET consists of a single dot
and two electrodes (source and drain). The mechanism of
interactions between two SETs mainly arises from the interdot
interactions such as interdot Coulomb repulsion and interdot
electron hopping. The latter is negligible as a result of high
$SiO_2$ barrier. Therefore, we take into account only the interdot
Coulomb repulsion in the calculation of current-voltage
characteristics. Kim and Hershfield studied the interdot spin-spin
interaction effect on the tunnelling current for the parallel
double QDs in the Kondo regime[10]. On the contrary, this interdot
spin-spin interaction is negligible for the Coulomb blockade
regime and therefore, it is not taken into account in this study.

\section{Energy levels and particle interactions}
To construct the Anderson model simulating the system shown in
Fig. 1, we calculate the energy levels, intradot and interdot
Coulomb interactions, and interdot electron hopping strengths
using the effective mass model. These physical parameters are
important in the calculation of current-voltage characteristics.
Two cylindrical Ge QDs are embedded into a $SiO_2$ layer with a
finite width W. The layer is then placed in contact with metals to
form an n-i-n structure. The metallic electrodes are considered
since they can avoid carriers frozen occurred in the electrodes at
low temperature[11]. To calculate the electronic structures of
QDs, we consider the cylindrical Ge QDs with a radius $R_0$ and
$h=2R_0$. The conduction band of Ge has four equivalent valleys
aligned along [1,1,1],[1,-1,-1],[-1,1,-1], and [-1,-1,1][12]. In
this calculation, the multi-valleys and image charge effects are
ignored for simplicity. Within the effective mass model, the
Hamiltonian is

\begin{eqnarray}
& &[-\nabla \frac {\hbar^2} {2m^*(\rho,z)} \nabla +
V_{QD,i}(\rho,z)] \psi_i({\bf r})\\
\nonumber &= & E_i \psi_i({\bf r}),
\end{eqnarray}

where ${m^{*}_e(\rho,z)}$ denotes the position-dependent electron
effective mass. Hence, $m_{S}^* = m_e$ for $SiO_2$ and
$m^{-1}_{Ge}=1/3 (1/m_{\ell}+2/m_t)$ for Ge,where
$m_{\ell}=1.59m_e$ and $m_t=0.0823m_e$ are the electron
longitudinal and transverse effective mass of bulk material.
$V_{QD,i}(\rho,z)$ is approximated by a constant potential $V_0 $
in the QD region. Its value is determined by the conduction band
offset between Ge and $SiO_2$. Besides the energy levels, the
Coulomb interactions should not be neglected for the electron
transport through small QDs. The Coulomb interactions are
calculated according to
\begin{equation}
 U_{i,j}= \int d{\bf r}_1 \int d{\bf r}_2 \frac {e^2 [n_i({\bf r}_1)n_j({\bf r}_2)]}
{\epsilon_0({\bf r_1};{\bf r_2}) |{\bf r}_{1}-{\bf r}_{2}|},
\end{equation}
where $n_i({\bf r_1})=|\psi_i(\rho,z,\phi)|^2$ is the particle
density of QD, and $\epsilon_0({\bf r}_1;{\bf r_2})$ denotes the
position-dependent static dielectric constant. For the purpose of
constructing approximate wave functions, we place the system in a
large cylindrical confining box with the length $L~$ and radius $R
~$ ($L$ and $R $ must be much larger than those of cylindrical Ge
QDs). Here $L = 60 nm $ and $R = 40 nm $ are adopted. We solve the
eigenfunctions of the effective-mass Hamiltonian by the Ritz
variational method. The wave functions are expanded in a set of
basis functions, which are chosen to be products of Bessel
functions and sine waves
$\psi_{n,\ell,m}(\rho,z,\phi)=J_{\ell}(\beta_n\rho)e^{i\ell \phi}
sin(k_m(z+L/2))$, where $k_m=m\pi/L,m=1,2,3..$, $J_{\ell}$ is the
Bessel function of the order of $\ell$ and $\beta_n R$ is the
$n$th zero of $J_{\ell}$. The same set of basis functions has been
used by Marzin and Bastard[13] to calculate the quantum confined
states in a conical QD. The expression of the matrix elements of
the Hamiltonian of Eq. (1) can be readily obtained. Forty sine
functions multiplied by fifteen Bessel function for each angular
function ($\ell$ = 0 or 1) are used to diagonalize the
Hamiltonian. Fig. 2 show the lowest two energy levels of a single
dot as a function of radius $R_0$ for $\ell=0$. Note that the
lowest energy level for $\ell=1$, not shown in Fig. 2, is higher
than the energy level denoted by the dashed line. To determine
physical parameters such as intradot Coulomb interactions, and
interdot Coulomb interactions, we consider the dot A of radius
$R_0=4.6 nm$ and the dot B of $R_0=4.4 nm$. The ground state
energies for the dot A and dot B are, respectively, $0.108 eV$ and
$0.118 eV$. Fig. 3 shows the intradot Coulomb interactions and
interdot interactions as a function of dot separation (ds), which
is defined as the distance between the bottom of dot A and dot B.
The solid line and dashed line denote the intradot Coulomb
interactions of the dot B and dot A ($U_B=37.18 meV$ and
$U_A=36.43meV$), the dotted line denotes the interdot Coulomb
interactions ($U_{AB}$). Obviously, the intradot Coulomb
interactions are almost independent of the dot separation. It is
also found that the interdot electron hopping energy, t, is
smaller than $0.1 meV$ for $ds \geq 16 nm$. This negligible t is
due to high potential barrier between Ge QDs and $SiO_2$. This
implies that the probability of charge transfer from the dot A to
the dot B is small.

\section{Tunnelling current}
Now, we construct the following Hamiltonian to describe the system
as shown in Fig. 1,

\begin{eqnarray}
H&=&\sum_{k,\sigma,\ell} \epsilon_k
a^{\dagger}_{k,\sigma,\ell}a_{k,\sigma,\ell}+\sum_{i=A,B;\sigma}
E_i d^{\dagger}_{i,\sigma} d_{i,\sigma} +\sum_{i,j,\sigma,\sigma'}
U_{i,j} n_{i,\sigma} n_{j,\sigma'}\\ \nonumber
&+&\sum_{k,\sigma,\ell,i}
V_{k,\ell,i}a^{\dagger}_{k,\sigma,\ell}d_{i,\sigma}+h.c
\end{eqnarray}
where $a^{\dagger}_{k,\sigma,\ell}$ creates an electron of
momentum $k$ and spin $\sigma$ with energy $\epsilon_k$ in the
$\ell$ electrode. $d^{\dagger}_{i,\sigma}$ creates an electron
inside the SET's QDs with orbital energy $E_i$, $U_{ij}$ describes
the intradot and interdot Coulomb interactions and $V_{k,\ell,i}$
describes the coupling between the band states and QDs.The
Hamiltonian given by Eq. (3) is based on the Anderson model with
two energy levels[14]. The time-independent tunnelling current
from the left leads (source electrodes) can be expressed as (ref.
[14]):

\begin{equation}
J=\frac{-2e}{\hbar}\sum_{i=A,B} \int \frac{d\epsilon}{2\pi}
\Gamma_{L,i}(\epsilon)Im[\frac{1}{2}G^{<}_{i,\sigma}(\epsilon)+f_{L,i}(\epsilon)G^{r}_{i,\sigma}(\epsilon)],
\end{equation}
where
$f_{\ell,i}(\epsilon)=(exp[(\epsilon-\mu_{\ell,i})/k_BT]+1)^{-1}$
is the Fermi distribution function of the $\ell=L,R$ lead.
$\Gamma_{\ell,i}(\epsilon)=2\pi \sum_k |V_{k,i}|^2
\delta(\epsilon-\epsilon_k)$ denotes the electron tunnelling rate
from the $i$th dot to the $\ell$ lead. Notations $e$ and $\hbar$
denote the electron charge and Plank's constant. The lesser
Green's function $G^{<}_{i,\sigma}(\epsilon)$ and retarded Green's
function $G^{r}_{i,\sigma}(\epsilon)$ are the Fourier
transformation of $G^{<}_{i,\sigma}(t)\equiv i\langle
d^{\dagger}_i(0)d_i(t)\rangle$ and
$G^r_{i,\sigma}(t)=-i\theta(t)\langle
\{d_{i,\sigma}(t),d^{\dagger}_{i,\sigma}(0)\}\rangle $, where
$\theta(t)$ is a step function, the curly brackets denote the
anti-commutator, and the braket  $\langle ...\rangle  $ represents
the thermal average. Using Dyson equation, we obtain the lesser
Green's function
\begin{equation}
ImG^{<}_{i,\sigma}(\epsilon)=-\frac{\Gamma_{L,i}(\epsilon)
f_{L,i}(\epsilon) +\Gamma_{R,i}(\epsilon)
f_{R,i}(\epsilon)}{\Gamma_{L,i}(\epsilon)+\Gamma_{R,i}(\epsilon)}
2Im[G^r_{i,\sigma}(\epsilon)].
\end{equation}
Substituting Eq. (5) into Eq. (4), we obtain the tunnelling
current as

\begin{eqnarray}
J&=&\frac{-2e}{\hbar}\sum_{i=A,B} \int
\frac{d\epsilon}{2\pi}\frac{\Gamma_{i,L}(\epsilon)
\Gamma_{i,R}(\epsilon)}
{\Gamma_{i,L}(\epsilon)+\Gamma_{i,R}(\epsilon)}\\ \nonumber &
&[f_{L,i}(\epsilon-\mu_{L,i})
-f_{R,i}(\epsilon-\mu_{R,i})]ImG^r_{i,\sigma}(\epsilon).
\end{eqnarray}

For simplicity, these tunnelling rates will be assumed energy-and
bias-independent. Therefore, the calculation of tunnelling current
is entirely determined by the spectral function
$A=ImG^r_{i,\sigma}(\epsilon)$, which is the imaginary part of the
retarded Green's function $G^r_{i,\sigma}(\epsilon)$. The
expression of retarded Green's function
$G^{r}_{i,\sigma}(\epsilon)$ can be obtained by the equation of
motion technique[15-18]. The lowest order coupling between the
electrodes and QDs will be considered in the calculation of
$G^r_{e,\sigma}(\epsilon)$. The equation of motion for
$G^{r}_{i,\sigma}(t)$ leads to
\begin{small}
\begin{equation}
(\epsilon-E_i+i\frac{\Gamma_{i}}{2})G^r_{i,\sigma}(\epsilon)=1+U_i
G^r_{ii}(\epsilon) + U_{ij}(
G^r_{ij,1}(\epsilon)+G^r_{ij2}(\epsilon)),
\end{equation}
\end{small}
where the two particle Green's functions $G^r_{ii}(\epsilon)$,
$G^r_{ij1}(\epsilon)$ and $G^r_{ij,2}(\epsilon)$ arise from the
particle correlation and satisfy
\begin{small}
\begin{equation}
(\epsilon-(E_i+U_i)+i\frac{\Gamma_i}{2})G^r_{ii}(\epsilon)
=N_{i,-\sigma}+U_{ij}(G^r_{iji1}(\epsilon)+ G^r_{iji2}(\epsilon)),
\end{equation}
\begin{equation}
(\epsilon-(E_i+U_{ij})+i\frac{\Gamma_i}{2})
G^{r}_{ij1}(\epsilon)=N_{j,\sigma} +U_{i}
G^r_{iji1}(\epsilon)+U_{ij} G^r_{ijj}(\epsilon),
\end{equation}
\end{small}
and
\begin{small}
\begin{equation}
(\epsilon-(E_i+U_{ij})+i\frac{\Gamma_i}{2}) G^r_{ij2}(\epsilon)=
N_{j,-\sigma} +U_{i}
G^r_{iji2}(\epsilon)+U_{ij}G^r_{ijj}(\epsilon).
\end{equation}
\end{small}
The notations $N_{i,-\sigma}$, $N_{j,\sigma}$ and $N_{j,-\sigma}$
are the electron occupation numbers for the $i$ dot. Note that $i$
is not equal to $j$ in Eqs. (7)-(10). Now the two-particle Green's
functions are coupled to the three-particle Green's functions
$G^r_{iji1}(\epsilon)=\langle{n_{i,-\sigma} n_{j,\sigma}
d_{i,\sigma},d^{\dagger}_{i,\sigma}}\rangle$,
$G^r_{iji2}(\epsilon)=\langle{n_{i,-\sigma} n_{j,-\sigma}
d_{i,\sigma},d^{\dagger}_{i,\sigma}}\rangle$, and
$G^r_{ijj}(\epsilon)=\langle{n_{j,-\sigma}n_{j,\sigma}
d_{i,\sigma},d^{\dagger}_{i,\sigma}}\rangle$. The equation of
motion of the three particle Green's functions will lead to
coupling with the four particle Green's functions, where the
hierarchy terminates. Thus, these three particle Green's functions
can be expressed in the following closed form
\begin{eqnarray}
G^r_{iji1}(\epsilon) =
N_{i,-\sigma}N_{j,\sigma}(\frac{1-N_{j,-\sigma}}
{\epsilon-(E_i+U_{i}+U_{ij})+i\frac{\Gamma_i}{2}}\nonumber \\
+\frac{N_{j,-\sigma}}
{\epsilon-(E_i+U_{i}+2U_{ij})+i\frac{\Gamma_i}{2}}),
\end{eqnarray}

\begin{eqnarray}
G^r_{iji2}(\epsilon)=N_{i,-\sigma}N_{j,-\sigma}(\frac{1-N_{j,\sigma}}
{\epsilon-(E_i+U_{i}+U_{ij})+i\frac{\Gamma_i}{2}} \nonumber \\
+ \frac{N_{j,\sigma}}
{\epsilon-(E_i+U_{j}+2U_{ij})+i\frac{\Gamma_i}{2}}),
\end{eqnarray}
and
\begin{eqnarray}
G^r_{ijj}(\epsilon)= N_{j,-\sigma} N_{j,\sigma}
(\frac{1-N_{i,-\sigma}}{\epsilon-(E_i+2U_{ij})+i\frac{\Gamma_i}{2}}\nonumber \\
+\frac{N_{i,-\sigma}}{\epsilon-(E_i+2U_{ij}+U_{i})+i\frac{\Gamma_i}{2}}).
\end{eqnarray}

Substituting Eqs. (11), (12) and (13) into Eqs (9) and (10), after
some algebras, the retarded Green's function $G^{r}_{i,\sigma}$ is
given by

\begin{small}
\begin{eqnarray}
&  & G^r_{i,\sigma}(\epsilon)=
(1-N_{i,-\sigma})\{\frac{1-(N_{j,\sigma}+N_{j,-\sigma})+N_{j,\sigma}N_{j,-\sigma}}
 {\epsilon - E_{i} +i \frac{\Gamma_{i}}{2} }\\ \nonumber &
+&\frac{N_{j,\sigma}+N_{j,-\sigma}-2
N_{j,\sigma}N_{j,-\sigma}}{\epsilon - E_{i} - U_{ij} +
i\frac{\Gamma_i}{2}} +\frac{N_{j,\sigma}N_{j,-\sigma} }{\epsilon -
E_{i} - 2U_{ij}+ i\frac{\Gamma_{i}}{2}}\}\\ \nonumber& + &
N_{i,-\sigma}\{\frac{1-(N_{j,\sigma}+N_{j,-\sigma})+N_{j,\sigma}N_{j,-\sigma}}
{\epsilon - E_{i}- U_{i} + i\frac{\Gamma_{i}}{2}} \\ \nonumber &
+&\frac{N_{j,\sigma}+N_{j,-\sigma}-2
N_{j,\sigma}N_{j,-\sigma}}{\epsilon- E_{i}- U_{i}-U_{ij} +
i\frac{\Gamma_{i}}{2}}+ \frac{N_{j,\sigma}N_{j,-\sigma}}{\epsilon
- E_{i}- U_{i}- 2U_{ij}+ i\frac{\Gamma_{i}}{2}} \}.
\end{eqnarray}
\end{small}
Note that we ignored the electron-electron correlation effect on
the tunnelling rates $\Gamma_{i}$ in Eq. (14). The approximation
used to obtain Eq. (14) is adequate for the Coulomb blockade but
not for the Kondon effect. In Eq. (14) there are two main
branches, one branch describes one-particle propagation process
and the other describes two-particle propagation process. In each
branch there are three channels with different probabilities. That
the total probability of these channels is one indicates the
satisfaction of sum rule $1=\int
(d\epsilon/\pi)|ImG^r_{i,\sigma}(\epsilon)|$. Refs. [16] and [17]
calculated the the tunnelling current of vertically coupled QD
including interdot Coulomb repulsion and employed the mean-field
scheme to solve the retarded Green's function, in which the
resonant energy levels are found to be dependent on the occupation
numbers of $N_i$ and $N_j$. However, the physical interpretation
for those resonant energy levels becomes ambiguity. In our
calculation, the poles of Eq. (14) do not depend on electron
occupation numbers, which determine the probability for each
channel. The electron occupation numbers of Eq. (14) can be solved
in a self-consistent way using
\begin{small}
\begin{eqnarray}
N_{i,\sigma} & = &\int
\frac{d\epsilon}{2\pi}ImG^{<}_{i,\sigma}(\epsilon)\\ \nonumber
 &= &-\int \frac{d\epsilon}{\pi}
\frac{\Gamma_{i,L} f_L(\epsilon)+\Gamma_{i,R}
f_R(\epsilon)}{\Gamma_{i,L}+\Gamma_{i,R}}
ImG^r_{i,\sigma}(\epsilon).
\end{eqnarray}
\end{small}
$N_{i,\sigma}$ is limited to the region of $0 \le N_{i,\sigma} \le
1 $. Electron occupation number $N_A$ ($N_B$), according to Eq.
(14), affects the tunnelling current through dot B (A). Eq. (14)
is main result of this work. Based on Eq. (14), we calculate the
current-voltage characteristic curves.

\section{Results and discussions}
In this section, we perform the detail numerical calculation of
I-V characteristics. For simplicity, we assume that tunnelling
rates of $\Gamma_1 = 0.2 meV$ and $\Gamma_2 = 0.2 meV$ are
bias-independent. Meanwhile the chemical potentials of four
electrodes are assumed to be $5 meV$ below the energy level $E_A$
for zero bias. The electron Coulomb interactions can be regarded
as bias-independent due to very strong confinement effect from the
barrier height of $V_0=3.5eV$. To determine the interdot Coulomb
repulsion, we consider the interdot separation distance $ds=16
nm$, which offers $U_{A,B}=11.06meV$. Substituting Eq. (14) into
Eq. (15), the electron occupation numbers $N_A$ and $N_B$ (for
fixed electron spin) could be solved in a self-consistent way.
Figure 4 shows $N_A$ and $N_B$ as a function of applied voltage
$V_a$ at zero temperature. We consider zero temperature through
out this article. In order to exhibit the interdot Coulomb
repulsion effect, the dashed lines for $U_{A,B}=0$ are also
plotted in Fig. 4. In the absence of $U_{A,B}$, the retarded
Green's function of Eq. (14) has only two poles
\[
G^r_{i,\sigma}(\epsilon)=\frac{1-N_{i,-\sigma}}{\epsilon-E_i+i\frac{\Gamma_i}{2}}
+\frac{N_{i,-\sigma}}{\epsilon-E_i-U_i+i\frac{\Gamma_i}{2}}.
\]
Therefore, the staircase behavior of $N_i$ is generated by the
on-site Coulomb interactions $U_i$. Owing to $\Gamma_L =\Gamma_R$,
$N_i$ reaches around 0.33 when the applied voltage crosses the
energy level $E_i$, but not sufficiently to overcome the charging
energy $U_i$. This is the well-known Coulomb blockade effect. When
the applied voltage overcomes the Coulomb blockade, $N_i$ reaches
to 0.5. These fractional occupation numbers are the typical
statistic feature for an open system.

For $U_{A,B} \neq 0$, $N_i$ shows a complicated behavior. We see
that the "switch-on" voltage $V_{a1}$ for B-SET is not changed in
the presence of tunnelling current of A-SET. Nevertheless, the
original occupation number $N_B=0.33$ (for $U_{A,B}=0$) is reduced
as the value of $N_B=0.22$. Once $N_B$ is finite, the probability
of resonant energy level $E_A$ is reduced. Consequently, the value
of $N_A$ is suppressed. When the applied voltage $V_a$ reaches the
resonant energy level $E_A+U_{AB}$, $N_A$ increases its value
again. As the applied voltage is modulated to $V_{a2}$, the value
of $N_B$ increases since the resonant level $E_B+U_{AB}$ is
opened. $N_A$ is slightly reduced at $V_{a2}$. The reduction of
$N_A$ is small at $V_{a2}$ because the one particle process for
$N_A$ is determined by the factor
$(1-N_{A,-\sigma})(1-N_{B,\sigma}N_{B,-\sigma})$, where
$N_{B,\sigma}N_{B,-\sigma}$ is around 0.09. When the applied
voltage is tuned to $V_{a}=42 mV$ for $N_A$ and $V_a=53 mV$ for
$N_B$, the two particle process is occurred for both $N_A$ and
$N_B$.The values of $N_A$ and $N_B$ that are larger than 0.33 are
manifested results of exhibiting two particle process. Obviously,
$N_A$ and $N_B$ are still suppressed by the interdot Coulomb
repulsion , even though the applied voltages overcome the on-site
Coulomb interactions.

Once the electron occupation numbers $N_A$ and $N_B$ are
determined, the tunnelling current can be calculated by
substituting Eq. (14) into Eq. (6). Figure. 5 exhibits the
tunnelling current and differential conductance (defined as
$dJ(V_a)/dV_a$) as a function of applied voltage $V_a$. Diagrams
(a) and (b) display the current through the dot A and dot B,
respectively. We see that the interdot Coulomb repulsion not only
creates new resonant channels for the source electrons
transferring into the drain electrode, but also yields negative
differential conductances. Results of Fig. 5 clearly exhibit the
suppression of tunnelling current by the interdot Coulomb
repulsion. The negative differential conductance through the dot
A(B) is generated by a fact that the applied voltage is not
sufficient to overcome the Coulomb repulsion from the charge
resided at the dot B (A).

Finally, we study the dot B situated at the condition of
$\Gamma_{B,L} \gg \Gamma_{B,R}$. From experimental point of view,
this is due to extremely asymmetrical barrier width. For the dot B
with the condition of $\Gamma_{B,L} \gg \Gamma_{B,R}$, the B-SET
behaves as a closed system. We assume that the dot B can be
manipulated as the following three charge states; the empty state
$|B,0 \rangle$, one electron state with spin up or down
$|B,\uparrow \rangle$ ($|B,\downarrow \rangle$) and two electron
states $|B,\uparrow, \downarrow \rangle$. Figure. 6 shows the
tunnelling current and differential conductance as a function of
applied voltage for various charge states of the dot B; solid line
denotes $|B,0 \rangle$, dashed line denotes $|B,\uparrow \rangle$
($|B,\downarrow \rangle$) and dotted line denotes $|B,\uparrow,
\downarrow \rangle$. The separation of the first three peaks
clearly displays the interdot Coulomb repulsion $U_{A,B}$. This
indicates that the "switch-on" voltage of A-SET depends on the
charge number of dot B. Obviously, the dot B situated in a closed
system will significantly affects the "switch-on" voltage of
A-SET. This also demonstrates that an SET is a highly sensitive
charge detector.

\section{Summary}
For parallel double SETs implemented by QDs embedded into $SiO_2$,
the interdot Coulomb repulsion is more important than the interdot
electron hopping as a result of high potential barrier. The
interdot Coulomb repulsion not only creates new resonant energy
levels but also produces negative differential conductances.
Nevertheless, note that the "switch-on" voltage of SETs is not
changed when both two SETs behave as an open system. If one of
SETs such as B-SET in this study behaves as a closed system, and
then the role of B-SET acts like a trap. The "switch-on" voltage
of A-SET will be dependent on the charges resided at the QD of
B-SET. This implies that the precisely control of barrier width in
the implementation of circuit integrated by SETs is very crucial.

%{\bf ACKNOWLEDGMENTS}
This work was supported by National Science Council of Republic of
China Contract No. NSC-93-2215-E-008-014 and NSC 93-2120-M-008-002

%\section{Appendix A}

\mbox{}

\newpage

{\bf Figure Captions}

Fig. 1: Schematic diagram for parallel double single electron
transistors.

Fig. 2: The lowest two energy levels of a cylindrical Ge/$SiO_2$
QD for $\ell=0$ as a function of dot radius $R_0$.

Fig. 3: Particle Coulomb interaction strengths as a function of
dot separation $ds$.

Fig. 4: Electron occupation numbers as a function of applied
voltage.

Fig. 5: Tunnelling current and differential conductance as a
function of applied voltage.

Fig. 6: Tunnelling current and differential conductance through
the dot A for various charge states of the dot B.

%Fig. 1: Energies of the bound states of a cylindrical Ge/$SiO_2$
%QD as a function of dot radius $R_0$.

%Fig. 2: $U_{11}$ , $U_{12}$  and $U_{13}$  as a function of dot
%radius $R_0$.

%\end{small}
\end{document}